\documentclass[12pt]{iopart}
\usepackage{graphicx}
\eqnobysec

\newcommand{\beq}{\begin{equation}}
\newcommand{\beqa}{\begin{eqnarray}}
\newcommand{\eeq}{\end{equation}}
\newcommand{\eeqa}{\end{eqnarray}}
\newcommand{\abs}[1]{\vert{#1}\vert}
\newcommand{\bigmean}[1]{\left\langle#1\right\rangle}
\newcommand{\cum}[1]{\langle\!\langle#1\rangle\!\rangle}
\newcommand{\cint}[1]{\mathrel{\mathop{\int}\limits_{#1}^{}}}
\newcommand{\cri}{{\rm cr}}
\renewcommand{\d}{{\rm d}}
\newcommand\dis[1]{\displaystyle#1}
\renewcommand{\e}{{\rm e}}
\newcommand{\frad}[2]{\displaystyle{\displaystyle#1\over\displaystyle#2}}
\newcommand{\half}{{\textstyle{\frac12}}}
\renewcommand{\i}{{\rm i}}
\renewcommand{\l}{\ell}
\newcommand{\mean}[1]{\langle#1\rangle}
\renewcommand{\min}{{\rm min}}
\newcommand{\prob}[1]{\mathop{{\rm Prob}}\left\{#1\right\}}
\newcommand{\w}{\widehat}
\renewcommand{\L}{\widetilde}
\newcommand{\DD}{{\cal D}}
\newcommand{\LL}{{\cal L}}
\newcommand{\MM}{{\cal M}}
\newcommand{\RR}{{\cal R}}
\newcommand{\TT}{{\cal T}}
\renewcommand{\Re}{\mathop{{\rm Re}}}

\begin{document}

\title{Statistics of quantum transmission in one dimension with broad disorder}

\author{D Boos\'e\dag\ and J M Luck\ddag}

\address{\dag\ Laboratoire de Physique Th\'eorique, UMR 7085 of CNRS
and Universit\'e Louis Pasteur, 3, rue de l'Universit\'e,
67084 Strasbourg Cedex, France}

\address{\ddag\ Service de Physique Th\'eorique, URA 2306 of CNRS,
CEA Saclay, 91191 Gif-sur-Yvette Cedex, France}

\begin{abstract}
We study the statistics of quantum transmission through a
one-dimen\-sio\-nal disordered system modelled by a sequence of
independent scattering units.
Each unit is characterized by its length and by its action, which is
proportional to the logarithm of the transmission probability through this unit.
Unit actions and lengths are independent random variables,
with a common distribution that is either narrow or broad.
This investigation is motivated by
results on disordered systems with non-stationary random potentials
whose fluctuations grow with distance.

In the statistical ensemble at fixed total sample length
four phases can be distinguished, according to the values of the indices
characterizing the distribution of the unit actions and lengths.
The sample action, which is
proportional to the logarithm of the conductance across the sample,
is found to obey a fluctuating scaling law,
and therefore to be non-self-averaging, in three of the four phases.
According to the values of the two above mentioned indices,
the sample action may typically grow less rapidly than linearly
with the sample length (underlocalization),
more rapidly than linearly (superlocalization),
or linearly but with non-trivial sample-to-sample fluctuations
(fluctuating localization).

\end{abstract}


\eads{\mailto{boose@lpt1.u-strasbg.fr},\mailto{jean-marc.luck@cea.fr}}

\maketitle

\section{Introduction}

The phenomenon of localization of a quantum particle
by a random potential is well understood (see, e.g.,~\cite{l1,l2}),
especially in the one-dimensional case (see, e.g.,~\cite{pen}).
It is known that this phenomenon has a major influence
on coherent quantum transport.
The amplitude $\TT$ of transmission through a sample is therefore a
quantity of central importance in this context.
The interpretation of experiments about the effects of localization
on transport is done with the help of the (two-probe)
Landauer formula~\cite{lan,il}, which relates the probability of
transmission $\abs{\TT}^2$ through a sample
to its conductance $g$ at zero temperature,
\beq
g=\frad{2e^2}{h}\,\abs{\TT}^2.
\label{lanfor}
\eeq

In the one-dimensional disordered systems usually considered, namely
those modelled by stationary random potentials with
short-range correlations, all the eigenstates are exponentially localized.
Thus any sample whose length $L$ is much greater than the
localization length $\xi$ is an insulator~\cite{l1,l2,pen}.
The probability of transmission
is a strongly fluctuating quantity in the insulating regime
and so it is preferable to consider its logarithm~\cite{four,aaa}.
It is therefore appropriate to introduce the sample action
\beq
S=-\frad{1}{2}\ln\abs{\TT}^2.
\label{sdef}
\eeq
The statistics of the transmission probability
in the insulating regime are determined by the distribution of $S$.
The average of the sample action over all the configurations
of the random potential grows asymptotically as
\beq
\mean{S}\approx\frad{L}{\xi}.
\eeq
The sample action $S$ is self-averaging in the strong sense that each
cumulant of its distribution increases
in direct proportion to the sample length~\cite{pen}.
Said otherwise, the distribution of $S$ is similar to that of the free energy
of a one-dimensional disordered thermodynamical system;
the transfer matrix formalism~\cite{pen,cpv,alea}
is well adapted to demonstrate this deep analogy.

The situation is quite different for one-dimensional disordered
systems modelled by non-stationary random potentials
whose fluctuations grow with distance~\cite{super1,super2,jmls}.
Typical eigenstates turn out to be superlocalized
rather than localized in samples whose length exceeds
a well-defined crossover length $\l_\cri$,
which diverges in the weak-disorder regime.
The action $S$ for long enough samples ($L\gg\l_\cri$)
exhibits two novel features with respect to
the usual one-dimensional localization problem:
it grows faster than linearly with $L$ and it is not self-averaging
in the sense that it keeps on fluctuating in arbitrarily large samples.
For instance, in the case of a self-affine Gaussian random potential
whose fluctuations increase as a power law with a Hurst exponent $H$
in the range $0<H<1$,
the scaling law of the sample action is the following~\cite{jmls}:
\beq
S\approx\left(\frad{E}{\l_\cri^H}\right)^{1/2}L^{1+H/2}\,\cal Y.
\label{shurst}
\eeq
Here $E$ denotes the energy of the particle,
$\l_\cri$ the above mentioned crossover length,
and~$\cal Y$ a rescaled random variable
whose distribution is a universal function
in the sense that it is determined by one parameter only,
namely the value of $H$.

In this work we address the question
whether the properties of the two previous
classes of models can be combined into a single model of
one-dimensional disordered system, in which the concomitant features
of self-averaging and conventional localization
hold in a certain range of parameters but are replaced
by the concomitant properties of lack of self-averaging and departure from
conventional localization in another range.
More specifically,
the purpose of this paper is to present and study a model which exhibits
the above properties
and is simple enough to be exactly solvable by analytical means.
The article is organized as follows.
In Section~2 we describe the model and set the notation used
throughout the paper.
In Section~3 we study the statistics of the transmission probability
through samples consisting of a large but fixed number $N$ of units.
In Section~4 we study the statistics of the transmission probability through
samples whose length $L$ is large but fixed.
The results and their physical consequences are discussed in Section~5.
Finally, an appendix is devoted to the derivation of the formula
of the sample action on which our calculations are based.

\section{The model}

\begin{figure}[htb]
\begin{center}
\includegraphics[angle=90,height=1.4truecm]{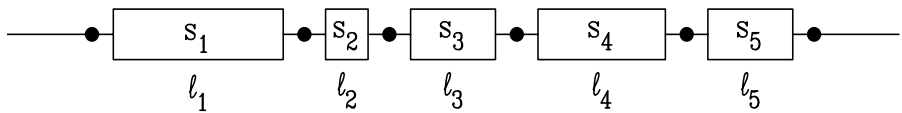}
\caption{\small
Schematic picture of the model considered in this paper,
consisting of a disordered array of elementary units
whose lengths $\l_n$ and actions $s_n$
are independent random variables.}
\label{fboxes}
\end{center}
\end{figure}

The model considered in this paper
consists of a disordered array of elementary units,
put together end to end on a line.
Each of these units consists of a disordered part
connected to two perfect leads, as shown pictorially in Figure~\ref{fboxes}
where disordered segments are represented by boxes, leads by line segments,
and the connection between each unit and the next one by a dot.

The $n$th unit has a length~$\l_n$
and its transport properties are characterized by the
transmission and reflection amplitudes $t_n$, $r_n$, and $r'_n$.
We shall work in the regime in which the probability of transmission
through every unit is very small ($\abs{t_n}^2\ll1$).
As explained in the appendix, in this weak transmission regime
and with the simplifying assumption that the reflection phases
are random and uniform, the law of addition for the actions~(\ref{adds}) holds.

The basic observables of our model are therefore the following.
For a sample made of $N$ units,
we consider its total action $S_N$ and total length $L_N$.
These quantities are simply expressed as sums,
\beq
S_N=\sum_{n=1}^N s_n,\quad L_N=\sum_{n=1}^N\l_n,
\label{sum}
\eeq
where the actions $s_n$ and lengths $\l_n$ of the units
are the microscopic variables of the model.
The simplicity of these definitions of the basic observables
makes the present model exactly solvable by analytical means.

Disorder is introduced into the model by assuming that the unit actions
$s_n$ and the unit lengths $\l_n$ are mutually independent random variables.
The actions $s_n$ are assumed to be greater than some minimal action $s_\min>0$
and to be distributed identically,
with a common continuous probability distribution
defined by the density $f_s(s)$.
Similarly, the lengths $\l_n$ are assumed
to be greater than some minimal length $\l_\min>0$
and to be distributed identically,
with a common distribution defined by the density $f_\ell(\ell)$.
Thus the sequences of random variables $S_N$ and $L_N$
obtained by adding units to a sample
constitute examples of a renewal process (see, e.g.,~\cite{renewal}).

The samples may be classified in two different ways,
either by their number $N$ of units, ignoring their length,
or by their length $L$, ignoring the number of units they consist of.
These two different points of view give rise to the following
two different statistical ensembles.

\begin{itemize}

\item {\it Statistical ensemble of the samples with a fixed number $N$ of
units}.
In this ensemble the sample action $S_N$ and the sample length $L_N$,
equation~(\ref{sum}),
are the sums of a {\it fixed} number~$N$ of independent random variables.

\item {\it Statistical ensemble of the samples with a fixed length $L$}.
In this ensemble the sample action, denoted by $S_L$, is
the sum of a $\it random$ number $N_L$ of independent random variables,
\beq
S_L=\sum_{n=1}^{N_L} s_n.
\label{sumlen}
\eeq
For a given configuration $\{\l_n\}$ of the unit lengths,
$N_L$ is defined as the number of units
which are {\it entirely} contained in the sample of fixed length $L$.
This choice is motivated by simplicity.
The random number $N_L$ therefore obeys the inequalities
\beq
L_{N_L}<L<L_{N_L+1},
\quad\hbox{i.e.,\ \ }\sum_{n=1}^{N_L}\l_n<L<\sum_{n=1}^{N_L+1}\l_n.
\label{nldef}
\eeq
Since the length of any unit is greater than the minimal length $\l_\min$,
the random integer~$N_L$ can take any value between 0 and
its maximal value $\lbrack L/\l_\min\rbrack$,
the symbol~$\lbrack x\rbrack$ denoting the integer part of the number $x$.

\end{itemize}

These two statistical ensembles will be carefully distinguished throughout
the following and successively investigated in Sections~3 and~4.
The statistics of the transmission may indeed exhibit qualitative differences
between both ensembles.

Generalizing a notation already introduced, we shall denote
the probability density of a random variable $\eta$ by $f_\eta(\eta)$.
The moment generating function and moment function of $\eta$
will be denoted by $\w f_\eta(x)$ and $m_\eta(p)$ respectively.
These functions are defined as
\beq
\w f_\eta(x)=\mean{\e^{-x\eta}}=\int\e^{-x\eta}\,f_\eta(\eta)\,\d\eta,\quad
m_\eta(p)=\mean{\eta^p}=\int\eta^p\,f_\eta(\eta)\,\d\eta,
\label{momdef}
\eeq
where the real parts of the complex variables $x$ and $p$
take their values in ranges such that the integrals are convergent.
The moment function is only defined for a positive random variable $\eta$.
The inverse formulae are
\beq
f_\eta(\eta)=\int\frad{\d x}{2\pi\i}\,\e^{x\eta}\,\w f_\eta(x)
=\int\frad{\d p}{2\pi\i}\,\eta^{-p-1}\,m_\eta(p).
\label{mominv}
\eeq
The moment generating function and the moment function are linked together
by the identity
\beq
m_\eta(-p)=\frad{1}{\Gamma(p)}\int_0^\infty\w f_\eta(x)\,x^{p-1}\,\d x
\quad(\Re p>0),
\label{mm}
\eeq
where the symbol $\Re z$ denotes the real part of the complex number $z$
and $\Gamma(z)$ the gamma function.
The identity~(\ref{mm}) is easily
verified by using the definition of $\w f_\eta(x)$, equation~(\ref{momdef}),
on the right-hand side and evaluating the integral in~$x$ with the help
of the first of the pair of identities~(\cite[p.~317]{gry})
\beq
\matrix{
\dis{\int_0^\infty x^{p-1}\,\e^{-xy}\,\d x=\frad{\Gamma(p)}{y^p}}\quad\hfill&
(\Re p>0,~\Re y>0),\hfill\cr
\dis{\cint{\Re y>0}
\frad{\d y}{2\pi\i}\,\frad{\e^{xy}}{y^p}=\frad{x^{p-1}}{\Gamma(p)}}\hfill&
(\Re x>0).\hfill}
\label{idens}
\eeq

It is convenient to characterize the probability densities
$f_s(s)$ and $f_\l(\l)$ by the exponents $\mu$, $\nu$
and the scales $s_0$, $\l_0$ of the power law fall off of the
corresponding complementary distribution functions:
\beq
\matrix{
\prob{s_1>s}=\dis{\int_s^\infty f_s(s_1)\,\d s_1}
\approx\left(\frad{s_0}{s}\right)^\mu\quad\hfill&(s\to\infty),\cr
\prob{\l_1>\l}=\dis{\int_\l^\infty f_\l(\l_1)\,\d\l_1}
\approx\left(\frad{\l_0}{\l}\right)^\nu\quad\hfill&(\l\to\infty).}
\label{foff}
\eeq
The notation $x\approx y$ used here means that the ratio $x/y$
tends to unity in the considered limit;
it will be used extensively in the following.
We shall also use the notation $x\sim y$, which has the weaker
meaning that $x$ and $y$ are only proportional in the considered limit,
up to an unimportant prefactor.

We shall consider the following classes of probability distributions
$f_s(s)$ and $f_\l(\l)$.

\begin{itemize}

\item {\it Narrow distributions ($\mu>2$; $\nu>2$)}.
These are the distributions whose first two moments at least are finite.
Thus we have
\beq
\matrix{
\w f_s(x)=1-x\mean{s}+\half x^2\mean{s^2}+\cdots,\hfill\cr
\w f_\l(x)=1-x\mean{\l}+\half x^2\mean{\l^2}+\cdots,\hfill}
\label{f2}
\eeq
where $\mean{s}$ and $\mean{s^2}$ (resp.~$\mean{\l}$ and $\mean{\l^2}$)
are the first two moments of $f_s(s)$ (resp.~$f_\l(\l)$).
Said otherwise, the narrow distributions are those for which the
central limit theorem holds.

The narrow distributions with a complementary
distribution function that falls off more rapidly than any power law
(we have then formally $\mu=\infty$ or $\nu=\infty$)
have the property that all their moments are finite, and so
\beq
\matrix{
\w f_s(x)=\dis{\sum_{k\ge0}}\frad{(-x)^k}{k!}\mean{s^k}
=\exp\left(\sum_{k\ge1}\frad{(-x)^k}{k!}\cum{s^k}\right),\hfill\cr
\w f_\l(x)=\dis{\sum_{k\ge0}}\frad{(-x)^k}{k!}\mean{\l^k}
=\exp\left(\sum_{k\ge1}\frad{(-x)^k}{k!}\cum{\l^k}\right),\hfill}
\label{fnarrow}
\eeq
where $\mean{s^k}$ and $\cum{s^k}$ (resp.~$\mean{\l^k}$ and $\cum{\l^k}$)
are the moment and cumulant of order~$k$ of $f_s(s)$ (resp.~$f_\l(\l)$).

\item {\it Broad distributions ($\mu<2$; $\nu<2$)}.
These are the distributions whose second moment is infinite.
In other words, the broad distributions are those for which the central
limit theorem does not hold.
Two cases have to be distinguished.
If $\mu<1$ (resp.~$\nu<1$), the first moment of $f_s(s)$ (resp.~$f_\l(\l)$)
is also infinite.
On the other hand, if $1<\mu<2$ (resp.~$1<\nu<2$), the first moment of
$f_s(s)$ (resp.~$f_\l(\l)$) is still finite.
The expansions of the moment generating functions as $x\to 0^+$ begin~as
\beq
\matrix{
\w f_s(x)=\left\{\matrix{
1-\Gamma(1-\mu)(s_0x)^\mu+\cdots\hfill&(\mu<1),\hfill\cr
1-\mean{s}x+\abs{\Gamma(1-\mu)}(s_0x)^\mu+\cdots\hfill\quad&(1<\mu<2),\cr}
\right.
\cr\cr
\w f_\l(x)=\left\{\matrix{
1-\Gamma(1-\nu)(\l_0x)^\nu+\cdots\hfill&(\nu<1),\hfill\cr
1-\mean{\l}x+\abs{\Gamma(1-\nu)}(\l_0x)^\nu+\cdots\hfill\quad&(1<\nu<2),\cr}
\right.
}
\label{fbroad}
\eeq
where the dots represent regular and singular terms of higher order~\cite{gl}.
Disorder modelled by broad distributions will be called broad disorder.

\end{itemize}

\section{Transmission statistics in the ensemble of samples with a fixed
number $N$ of units}

The statistics of the transmission probability through samples
with a fixed number~$N$ of units are determined by the distribution of
the sample action $S_N$, equation~(\ref{sum}).
We shall be interested here in the scaling properties of $S_N$
in the large $N$ limit.
This study may be considered as an introduction to that done in Section~4,
in which the most interesting results are derived.

It is useful to introduce the
moment generating function $\w f_{S_N}(x)$ of $S_N$.
Since the unit actions are mutually independent random variables with a common
distribution, $\w f_{S_N}(x)$ takes the form
\beq
\w f_{S_N}(x)=\w f_s(x)^N.
\label{fln}
\eeq
The scaling law of $S_N$ can be obtained from the expression of
this quantity in the large~$N$ and small~$x$ limits.
Three cases have to be distinguished,
as the expansions~(\ref{f2}) and~(\ref{fbroad}) of $\w f_s(x)$ for small $x$
are different for $\mu>2$, $\mu<1$, and $1<\mu<2$.
We shall discuss these three cases below in this order.
The results of this section also apply to the
distribution of the sample length $L_N$,
provided we replace the index~$\mu$ and the parameters $\mean{s}$ and $s_0$
in the obtained results
by the index~$\nu$ and the parameters $\mean{\l}$ and~$\l_0$ respectively.

\subsection{Self-averaging with normal fluctuations: $\mu>2$}

This is the easiest case to study,
as the distribution of the unit actions $s_n$ is narrow.
Using the expression~(\ref{f2}) of $\w f_s(x)$ in~(\ref{fln}), we obtain
\beq
\w f_{S_N}(x)\approx\exp\left(N\left(-x\mean{s}+\half x^2\cum{s^2}\right)\right)
\eeq
in the relevant regime ($N$ large and $x$ small).
Setting
\beq
S_N\approx N\mean{s}+(N\cum{s^2})^{1/2}\,\eta,
\label{n1}
\eeq
we find that the moment generating function of the rescaled
random variable~$\eta$ is
\beq
\w f_\eta(y)=\e^{y^2/2}.
\eeq

Equation~(\ref{n1}) gives the scaling law of the sample action
in the present case.
The fluctuations of $S_N$ about its mean $N\mean{s}$ increase as $N^{1/2}$,
and so become negligible in relative value in the large $N$ limit.
Hence $S_N$ is self-averaging.
The distribution of $\eta$ is Gaussian,
\beq
f_\eta(\eta)=\frad{\e^{-\eta^2/2}}{\sqrt{2\pi}}.
\eeq

In the case of narrow distributions with a
complementary distribution function that falls off more rapidly than any power
law, $S_N$ is self-averaging in the strong sense that each
cumulant of its distribution is exactly proportional to $N$:
\beq
\cum{S_N^k}=N\cum{s^k}.
\eeq
This can be checked by using the expression~(\ref{fnarrow})
of $\w f_s(x)$ in~(\ref{fln}).

\subsection{Fluctuating scaling law: $\mu<1$}

In this case the distribution of the unit actions $s_n$
is broad, with an infinite mean $\mean{s}$.
Using the first expression~(\ref{fbroad}) of $\w f_s(x)$
in~(\ref{fln}), we obtain
\beq
\w f_{S_N}(x)\approx\exp\left(-N\Gamma(1-\mu)(s_0x)^\mu\right)
\label{wn2}
\eeq
in the relevant regime ($N$ large and $x$ small).
Setting
\beq
S_N\approx s_0(\Gamma(1-\mu))^{1/\mu}N^{1/\mu}\,\LL_\mu,
\label{n2}
\eeq
we find that the positive rescaled random variable $\LL_\mu$
has a non-trivial limiting distribution, whose moment generating function is
\beq
\w f_{\LL_\mu}(y)=\e^{-y^\mu}.
\label{n2e}
\eeq

Equation~(\ref{n2}) gives the scaling law of the sample action
in the present case.
It shows that $S_N$ increases as $N^{1/\mu}$,
that is, faster than the number $N$ of units,
because the exponent $1/\mu$ is greater than unity.
Furthermore, $S_N$ is not self-averaging
but keeps on fluctuating in the large $N$ limit.
The result~(\ref{n2}) therefore appears as a {\it fluctuating scaling law}.
The expression of the distribution of the rescaled random variable $\LL_\mu$
is obtained by using~(\ref{n2e}) in the first of the formulae~(\ref{mominv}),
\beq
f_{\LL_\mu}(\LL)=\cint{\Re y>0}\frad{\d y}{2\pi\i}\,\e^{y\LL-y^\mu}.
\label{flevy}
\eeq
This probability density is referred to as the (properly normalized)
L\'evy law of index~$\mu$.
It is a non-trivial universal function
in the sense that it is determined by one parameter only,
namely the value of $\mu$.
References~\cite{levy1} provide a comprehensive presentation of L\'evy laws,
and References~\cite{levy2} provide overviews of applications
of L\'evy variables in Physics.

Expanding $\e^{-y^\mu}$ as a power series in the integrand of~(\ref{flevy})
and integrating term by term with the help of the second
of the identities~(\ref{idens}), we obtain the convergent expansion
\beq
f_{\LL_\mu}(\LL)=\sum_{k\ge1}\frad{(-1)^k}{k!\,\Gamma(-k\mu)}\,\LL^{-(1+k\mu)}.
\eeq
Using the complement formula for the gamma function,
we find the alternative form
\beq
f_{\LL_\mu}(\LL)=\sum_{k\ge1}\frad{(-1)^{k-1}}{k!}\,\frad{\sin(k\pi\mu)}{\pi}
\,\Gamma(1+k\mu)\,\LL^{-(1+k\mu)}.
\label{levyinf}
\eeq
The behaviour of the probability density $f_{\LL_\mu}(\LL)$
at large values of $\LL$ is
determined by the leading order term of this expansion,
\beq
f_{\LL_\mu}(\LL)\approx\frad{\sin\pi\mu}{\pi}\,\Gamma(1+\mu)\,\LL^{-(1+\mu)}
\quad(\LL\to\infty).
\label{larl}
\eeq
Its behaviour at small values of $\LL$ is obtained from~(\ref{flevy})
by means of the method of steepest descent.
We thus find an exponentially fast fall off:
\beq
f_{\LL_\mu}(\LL)
\sim\exp\left(-(1-\mu)\left(\frad{\mu}{\LL}\right)^{\mu/(1-\mu)}\right)
\quad(\LL\to0).
\label{levy0}
\eeq
The density $f_{\LL_\mu}(\LL)$ admits a particularly simple
expression in closed form for $\mu=1/2$, which is
\beq
f_{\LL_{1/2}}(\LL)=\frad{\e^{-1/(4\LL)}}{2\sqrt{\pi\LL^3}}.
\eeq

The expression of the moment function $m_{\LL_\mu}(p)$ may be derived
starting from~(\ref{mm}).
Using~(\ref{n2e}) in the right-hand side of the identity, we obtain it in three
steps.
We first evaluate the integral with the help
of the relation~(\cite[p.~342]{gry})
\beq
\int_0^\infty x^{p-1}\e^{-x^{\mu}}\,\d x=\frad{1}{\mu}\Gamma(p/\mu)
\quad(\mu>0,~\Re p>0).
\eeq
We then rewrite the result using the difference formula for the gamma function.
We finally extend the obtained result to negative values of $\Re p$ by
analytic continuation.
We thus find the simple formula
\beq
m_{\LL_\mu}(p)=\frad{\Gamma(1-p/\mu)}{\Gamma(1-p)}\quad(\Re p<\mu).
\label{xmom}
\eeq

\subsection{Self-averaging with anomalous fluctuations: $1<\mu<2$}

In this case the distribution of the unit actions $s_n$ is
still broad, but it has a finite mean~$\mean{s}$.
Using the second expression~(\ref{fbroad}) of $\w f_s(x)$ in~(\ref{fln}), we
obtain
\beq
\w f_{S_N}(x)\approx
\exp\Bigl(N\left(-\mean{s}x+\abs{\Gamma(1-\mu)}\,(s_0x)^\mu\right)\Bigr)
\eeq
in the relevant regime ($N$ large and $x$ small).
Setting
\beq
S_N\approx N\mean{s}+s_0\abs{\Gamma(1-\mu)}^{1/\mu}N^{1/\mu}\,\LL_\mu,
\label{n3}
\eeq
we find that the moment generating function of the rescaled random variable
$\LL_\mu$ is now
\beq
\w f_{\LL_\mu}(y)=\e^{y^\mu}.
\label{n3e}
\eeq

Equation~(\ref{n3}) gives the scaling law of the sample action
in the present case.
This expression involves
two different microscopic characteristics of the distribution $f_s(s)$,
namely the mean $\mean{s}$ and the amplitude $s_0$
of the power law tail~(\ref{foff}).
The fluctuations of $S_N$ about its mean $N\mean{s}$ increase
as $N^{1/\mu}$, with $1/2<1/\mu<1$.
On the one hand, these fluctuations are much smaller than the mean action
in the large $N$ limit, and so $S_N$ is self-averaging.
On the other hand, these fluctuations increase more rapidly than $N^{1/2}$,
the power law of the central limit theorem
involved in the expression~(\ref{n1}).
Furthermore,
the distribution of the rescaled random variable $\LL_\mu$ is not Gaussian.

Using~(\ref{n3e}), we obtain that $\LL_\mu$ is distributed according to
the following L\'evy law of index $\mu$:
\beq
f_{\LL_\mu}(\LL)=\cint{\Re y>0}\frad{\d y}{2\pi\i}\,\e^{y\LL+y^\mu}.
\label{flevy1}
\eeq
This probability density is universal,
as it only depends on the value of the index $\mu$.
It is non-vanishing for all values of $\LL$.
Its fall off at large positive values of~$\LL$ can be derived
by linearizing~(\ref{flevy1}) as
\beq
f_{\LL_\mu}(\LL)\approx\cint{\Re y>0}\frad{\d y}{2\pi\i}\,\e^{y\LL}\,y^\mu.
\eeq
Using again the second of the identities~(\ref{idens}), we thus obtain
\beq
f_{\LL_\mu}(\LL)\approx
\frad{\abs{\sin\pi\mu}}{\pi}\,\Gamma(1+\mu)\,\LL^{-(1+\mu)}
\quad(\LL\to+\infty).
\eeq
This expression is the first term of an expansion analogous to~(\ref{levyinf}).
This expansion is, however, asymptotic but divergent in the present case.
The behaviour of $f_{\LL_\mu}(\LL)$
at large negative values of $\LL$ is obtained from~(\ref{flevy1})
by means of the method of steepest descent.
We thus find
\beq
f_{\LL_\mu}(\LL)\sim
\exp\left(-(\mu-1)\left(\frad{\abs{\LL}}{\mu}\right)^{\mu/(\mu-1)}\right)
\quad(\LL\to-\infty).
\eeq
Since the exponent $\mu/(\mu-1)$ is larger than 2,
this fall off is of a superexponential type.

\section{Transmission statistics in the ensemble of samples with a fixed
length~$L$}

The statistics of the transmission probability through samples
with a fixed length~$L$ are determined by the distribution of
the sample action $S_L$, equation~(\ref{sumlen}).
The random character of $S_L$ has two origins,
as it is the sum of a random number $N_L$ of independent random variables.
We shall be interested here in the scaling properties of $S_L$
in the limit of large values of $L$.

It is again useful to introduce the moment generating function of
the sample action.
Its expression is now
\beq
\w f_{S_L}(x,L)=\sum_{n\ge0}p_n(L)\,\w f_{S_n}(x),
\eeq
where $p_n(L)=\prob{N_L=n}$ is the probability that a sample of length $L$
consists of exactly $n$ units
and $\w f_{S_n}(x)$ the moment generating function corresponding
to this particular number of units.
Using~(\ref{fln}), we obtain
\beq
\w f_{S_L}(x,L)=\sum_{n\ge0}p_n(L)\,\w f_s(x)^n.
\label{mgsl}
\eeq
When dealing with functions $f(L)$ that depend on the sample length,
it will prove useful to use their Laplace transform $\L{f}(u)$
with respect to this variable,
\beq
\L{f}(u)=\int_0^\infty\e^{-uL}\,f(L)\,\d L.
\eeq
Equation~(\ref{mgsl}) becomes
\beq
\L{\w f_{S_L}}(x,u)=\sum_{n\ge0}\L{p}_n(u)\,\w f_s(x)^n.
\label{ltmgf}
\eeq
The expression of the Laplace transform $\L{p}_n(u)$ can be obtained as follows.
The definition~(\ref{nldef}) of $N_L$ implies
\beqa
p_n(L)&=&\prob{L_n<L<L_{n+1}}\nonumber\\
&=&\prob{L>L_n}-\prob{L>L_{n+1}}.
\label{probdif}
\eeqa
The Laplace transform of the first of these probabilities is simply
\beqa
\int_0^\infty\prob{L>L_n}\,\e^{-uL}\,\d L
&=&\bigmean{\int_{L_n}^\infty\e^{-uL}\,\d L}\nonumber\\
&=&\frad{\mean{\e^{-uL_n}}}{u}
=\frad{\w f_{L_n}(u)}{u}
=\frad{\w f_\l(u)^n}{u},
\label{probn}
\eeqa
where the last equality has been obtained in analogy
with the derivation of~(\ref{fln}).
The Laplace transform of the second probability
is obtained upon replacing $n$ by $n+1$ in~(\ref{probn}).
It follows that (see, e.g.,~\cite{gl})
\beq
\L{p}_n(u)=\w f_\l(u)^n\,\frad{1-\w f_\l(u)}{u}\qquad(n\ge0).
\label{lappn}
\eeq

Using~(\ref{lappn}) in~(\ref{ltmgf}), we find that the Laplace transform of the
moment generating function has the following closed form expression:
\beq
\L{\w f_{S_L}}(x,u)=\frad{1-\w f_\l(u)}{u(1-\w f_\l(u)\w f_s(x))}.
\label{mw}
\eeq
This equation is analogous to
the Montroll-Weiss equation~\cite{mw}, which is well known in
the theory of the continuous time random walks (see, e.g.,~\cite{ctrw}).

In the following, the scaling law of $S_L$ will be derived by using~(\ref{mw})
in the small~$x$ and small $u$ limits.
The expressions~(\ref{f2}) and~(\ref{fbroad})
of $\w f_s(x)$ are different for $\mu>1$ and $\mu<1$,
as well as those of $\w f_\l(u)$ for $\nu>1$ and $\nu<1$.
This leads to four different phases, which are identified by
the conditions ($\mu>1$, $\nu>1$), ($\mu<1$, $\nu>1$), ($\mu>1$, $\nu<1$),
and ($\mu<1$, $\nu<1$).
We shall discuss these four phases in this order.
For the sake of simplicity
we shall not distinguish the situations of normal and anomalous
fluctuations when the sums are self-averaging.
Said otherwise, we shall not discuss separately
the cases $\mu>2$ and $1<\mu<2$, nor the cases $\nu>2$ and $1<\nu<2$.

\subsection{Phase~I: $\mu>1$ and $\nu>1$}

In this phase both means $\mean{s}$ and $\mean{\l}$ are finite.
The result~(\ref{n1}) implies that $S_N\approx N\mean{s}$
and similarly that $L_N\approx N\mean{\l}$,
up to fluctuations that become negligible in relative value
in the large $N$ limit.
Eliminating $N$ between both estimates, we obtain
\beq
S_L\approx\frad{\mean{s}}{\mean{\l}}\,L,
\label{l1}
\eeq
again up to fluctuations that become negligible in relative value at large $L$.
In order to obtain more accurate estimates
of the mean $\mean{S_L}$ and the fluctuations about it,
more restrictive conditions on the distributions $f_\l$ and $f_s$ are needed.

Let us consider first the case of narrow distributions
with $\mu>2$ and $\nu>2$.
Using the expressions~(\ref{f2}) of $\w f_s(x)$ and $\w f_\l(u)$ in~(\ref{mw})
and expanding this formula to second order in $x$ and $u$, we find
that the mean and variance of the distribution of the sample action
have the following expressions:
\beqa
&&\mean{S_L}\approx\frad{\mean{s}}{\mean{\l}}\left(L
+\frad{\cum{\l^2}-\mean{\l}^2}{2\mean{\l}}\right),\label{l1s1}\\
&&\cum{S_L^2}\approx\frad{\mean{\l}^2\cum{s^2}+\mean{s}^2\cum{\l^2}}
{\mean{\l}^3} L.
\label{l1s2}
\eeqa
Equation~(\ref{l1s1}) shows that the actual expression of the mean is not
exactly the one given by~(\ref{l1}).
There is a finite correction term which is proportional to
$\cum{\l^2}-\mean{\l}^2=\mean{\l^2}-2\mean{\l}^2$
and so may be of either sign.
Equation~(\ref{l1s2}) indicates that the variance increases
in direct proportion to $L$.
The first correction to this leading order result,
which has not been written explicitly, is a constant term for $\nu>3$
but grows as $\mean{s}^2L^{3-\nu}$ for $2<\nu<3$.

We also consider the case of narrow distributions with a
complementary distribution function that falls off
more rapidly than any power law.
In this case $S_L$ is self-averaging
in the strong sense that each cumulant of its distribution
increases in direct proportion to $L$:
\beq
\cum{S_L^k}\approx a_k L.
\label{cuml}
\eeq
The expressions of the coefficients $a_k$ can be obtained as follows.
Equation~(\ref{cuml}) means that the moment generating function
$\w f_{S_L}(x,L)$ has the following exponential growth
for large values of the sample length:
\beq
\w f_{S_L}(x,L)\sim\e^{A(x)L}.
\label{cumf}
\eeq
Here $A(x)$ is the generating function of the coefficients $a_k$,
\beq
A(x)=\sum_{k\ge1}\frad{(-x)^k}{k!}\,a_k.
\eeq
Equation~(\ref{cumf}) in turn
implies that the Laplace transform $\L{\w f_{S_L}}(x,u)$
has a simple pole of the form
\beq
\L{\w f_{S_L}}(x,u)\sim\frad{1}{u-A(x)}.
\eeq
On the other hand,~(\ref{mw}) shows that $\L{\w f_{S_L}}(x,u)$ has poles
whenever $\w f_\l(u)\w f_s(x)=1$.
Hence the generating function $A(x)$ is determined by the implicit equation
\beq
\w f_\l(A(x))\w f_s(x)=1.
\eeq
Using the expression~(\ref{fnarrow}) of $\w f_s(x)$
and $\w f_\l(x)$ in this equation and expanding it as a
power series in $x$, we can derive the expressions of the coefficients $a_k$
in a recursive way.
We find
\beq
\matrix{
a_1=\frad{\mean{s}}{\mean{\l}},\hfill\cr
a_2=\frad{\mean{\l}^2\cum{s^2}+\mean{s}^2\cum{\l^2}}{\mean{\l}^3},\hfill\cr
a_3=\frad{\mean{\l}^4\cum{s^3}-\mean{\l}\mean{s}^3\cum{\l^3}
+3\mean{\l}^2\mean{s}\cum{\l^2}\cum{s^2}+3\mean{s}^3\cum{\l^2}^2}{\mean{\l}^5},
}
\eeq
and so on.
As expected, the expression of the coefficient $a_1$ (resp.~$a_2$) agrees with
the one obtained from~(\ref{l1s1}) (resp.~(\ref{l1s2})).

\subsection{Phase~II: $\mu<1$ and $\nu>1$}

In this phase the mean $\mean{\l}$ is finite,
but the mean $\mean{s}$ is infinite.
The fact that $\mean{\l}$ is finite implies that the number of units
increases as $N\approx L/\mean{\l}$, by analogy with~(\ref{n1}).
We therefore surmise that the scaling law of $S_L$ is similar to the
one of $S_N$ in the case $\mu<1$.

The scaling law of $S_L$ can be obtained as follows.
Using the first expression~(\ref{fbroad}) of $\w f_s(x)$ and the
expression~(\ref{f2}) of $\w f_\l(u)$ in~(\ref{mw}), we obtain
\beq
\L{\w f_{S_L}}(x,u)\approx\left(u+\frad{\Gamma(1-\mu)(s_0x)^\mu}{\mean{\ell}}
\right)^{-1}
\eeq
in the relevant regime ($x$ and $u$ small).
The calculation of the inverse Laplace transform of $\L{\w f_{S_L}}(x,u)$
gives
\beq
\w f_{S_L}(x,L)
\approx\exp\left(-\Gamma(1-\mu)(s_0x)^\mu\frad{L}{\mean{\ell}}\right).
\eeq
We note that this expression becomes identical to~(\ref{wn2})
if we replace the ratio $L/\mean{\l}$ by the number $N$ of units.
This allows us to conclude that the scaling law of $S_L$ in Phase~II~is
\beq
S_L\approx s_0\left(\Gamma(1-\mu)\right)^{1/\mu}
\left(\frad{L}{\mean{\ell}}\right)^{1/\mu}\,\LL_\mu,
\label{l2}
\eeq
where the rescaled random variable $\LL_\mu$ is
distributed according to the L\'evy law, equation~(\ref{flevy}).
This result agrees with our surmise.

\subsection{Phase~III: $\mu>1$ and $\nu<1$}

In this phase the mean $\mean{s}$ is finite but the mean
$\mean{\l}$ is infinite.
The fact that $\mean{s}$ is finite implies that $S_N\approx N\mean{s}$, by
virtue of~(\ref{n1}).
We therefore surmise that the fluctuations of
$S_L$ are similar here to those of the number $N_L$ of units in the samples.

Following our surmise, we begin by deriving
the scaling form of the expression of $p_n(L)$
in the limit of large values of $L$.
Using the first expression~(\ref{fbroad}) of $\w f_\l(u)$
in~(\ref{lappn}), we obtain
\beq
\L{p}_n(u)\approx\Gamma(1-\nu)\l_0^\nu u^{\nu-1}
\,\exp\Bigl(-\Gamma(1-\nu)(\l_0u)^\nu n\Bigr)
\label{lpn}
\eeq
in the relevant regime ($n$ large, $u$ small), and so
\beq
p_n(L)\approx\Gamma(1-\nu)\l_0^\nu
\cint{\Re u>0}\frad{\d u}{2\pi\i}\,u^{\nu-1}
\,\exp\Bigl(-\Gamma(1-\nu)(\l_0u)^\nu n+uL\Bigr).
\label{pn}
\eeq

Let us introduce a rescaled variable $Z_\nu$ by setting
\beq
n\approx\frad{1}{\Gamma(1-\nu)}\left(\frad{L}{\l_0}\right)^\nu Z_\nu.
\label{ln3}
\eeq
This positive random variable has a limiting distribution
which is easily obtained from~(\ref{pn}).
Setting $z=uL$, we find
\beq
f_{Z_\nu}(Z)=\cint{\Re z>0}\frad{\d z}{2\pi\i}\,z^{\nu-1}\,\e^{z-Z z^\nu}.
\label{fz}
\eeq
This probability density admits an expression in closed
form for $\nu=1/2$, case in which it is a half-Gaussian:
\beq
f_{Z_{1/2}}(Z)=\frad{\e^{-Z^2/4}}{\sqrt{\pi}}\quad(Z>0).
\eeq

The random variable $Z_\nu$ can be expressed as
\beq
Z_\nu={\LL_\nu}^{-\nu},
\label{zxi}
\eeq
where $\LL_\nu$ is distributed according to
the L\'evy law, equation~(\ref{flevy}).
Indeed, setting $Z=\LL^{-\nu}$ and $z=x\LL$ in~(\ref{fz}), we
obtain the following expression for the probability density of $\LL_\nu$:
\beq
f_{\LL_\nu}(\LL)=\frad{\nu}{\LL}\cint{\Re x>0}\frad{\d x}{2\pi\i}
\,x^{\nu-1}\,\e^{x\LL-x^\nu}.
\eeq
This expression is then put into the canonical form~(\ref{flevy})
by means of an integration by parts.

The expression of the moment function of $Z_\nu$ is readily obtained by
means of~(\ref{xmom}) and~(\ref{zxi}).
We find
\beq
m_{Z_\nu}(p)=\frad{\Gamma(1+p)}{\Gamma(1+p\nu)}\quad(\Re p>-1).
\label{zmom}
\eeq
This result implies that the distribution of $Z_\nu$ is narrow
in the strong sense that all its moments are finite.
Setting $p=k$ in~(\ref{zmom}), we indeed obtain
\beq
\mean{Z_\nu^k}=\frad{k!}{\Gamma(1+k\nu)}
\eeq
for any integer $k$.

The relations~(\ref{ln3}) and~(\ref{zxi}) may also
be found in the following heuristic way.
By analogy with~(\ref{n2}), the
sample length $L_N$ has the following fluctuating scaling law in the case
$\nu<1$:
\beq
L_N=\l_0(\Gamma(1-\nu))^{1/\nu}\,N^{1/\nu}\,\LL_\nu.
\label{ls}
\eeq
Expressing $N$ in this equation in terms of the other quantities,
we obtain both above relations at once.

Let us now turn to the scaling law of $S_L$.
Using the first expression~(\ref{fbroad}) of $\w f_\l(u)$
and the expression~(\ref{f2}) of $\w f_s(x)$ in~(\ref{mw}), we obtain
\beq
\L{\w f_{S_L}}(x,u)\approx\frad{\Gamma(1-\nu)\l_0^\nu u^{\nu-1}}
{x\mean{s}+\Gamma(1-\nu)(\l_0u)^\nu}
\eeq
in the relevant regime ($x$ and $u$ small), and so
\beq
\L{f}_{S_L}(S_L,u)\approx\frad{\Gamma(1-\nu)\l_0^\nu u^{\nu-1}}{\mean{s}}
\,\exp\left(-\Gamma(1-\nu)(\l_0u)^\nu\frad{S_L}{\mean{s}}\right).
\eeq
We note that this expression becomes identical to~(\ref{lpn})
if we replace the ratio $S_L/\mean{s}$ by the number $n$ of units.
This allows us to conclude that the scaling law of $S_L$ in Phase~III is
\beq
S_L\approx\frad{\mean{s}}{\Gamma(1-\nu)}\left(\frad{L}{\l_0}\right)^\nu Z_\nu.
\label{s3}
\eeq
Our surmise has proved correct, as the same random variable $Z_\nu$,
equation~(\ref{zxi}),
enters the scaling laws of the number of units and of the sample action.

\subsection{Phase~IV: $\mu<1$ and $\nu<1$}

This last phase is the most interesting one,
because neither the mean $\mean{s}$ nor the mean~$\mean{\l}$ is finite.
This implies that $S_N\sim N^{1/\mu}$ and $L_N\sim N^{1/\nu}$,
by virtue of and analogy with~(\ref{n2}) respectively.
The combination of these two scaling laws gives $S_N\sim L_N^{\nu/\mu}$.
We therefore surmise that the sample action follows a fluctuating scaling law
of the form $S_L\sim L^{\nu/\mu}$ in this phase.

To obtain the scaling law of $S_L$, it is easier here to work with the
Laplace transform $\L{m}_{S_L}(p,u)$
of the moment function than with the one of the moment generating function.
These two functions are linked together by the identity
\beq
\L{m}_{S_L}(-p,u)=\frad{1}{\Gamma(p)}\int_0^\infty\L{\w
f_{S_L}}(x,u)\,x^{p-1}\,\d x\quad(\Re p>0),
\label{mmm}
\eeq
which is readily obtained by taking the Laplace transform of~(\ref{mm})
with respect to $L$.
Using the first expression~(\ref{fbroad}) of $\w f_s(x)$
and of $\w f_\l(u)$ in~(\ref{mw}), we find
\beq
\L{\w f_{S_L}}(x,u)\approx\frad{\Gamma(1-\nu)\l_0^\nu u^{\nu-1}}
{\Gamma(1-\nu)(\l_0u)^\nu+\Gamma(1-\mu)(s_0x)^\mu}
\label{wf4}
\eeq
in the relevant regime ($x$ and $u$ small).
Putting this expression into~(\ref{mmm}), we obtain
that of $\L{m}_{S_L}(p,u)$ by following the same
three-step procedure as in the derivation of the expression of the
moment function $m_{\LL_\mu}(p)$, equation~(\ref{xmom}), the integral
being evaluated with the help of the relation~(\cite[p.~292]{gry})
\beq
\int_0^\infty\frad{x^{p-1}}{1+zx^{\mu}}\,\d x
=\frad{\Gamma(p/\mu)\Gamma(1-{p/\mu})}{\mu\,z^{p/\mu}}\quad(0<\Re p<\mu).
\eeq
We thus find
\beq
\L{m}_{S_L}(p,u)=s_0^p
\left(\frad{\Gamma(1-\mu)}{\Gamma(1-\nu)}\right)^{p/\mu}
\frad{\Gamma(1-p/\mu)\Gamma(1+p/\mu)}{\Gamma(1-p)(\l_0u)^{p\nu/\mu}\,u}.
\label{comp1}
\eeq
The inverse Laplace transform is straightforward, and leads to
\beq
m_{S_L}(p,L)\approx s_0^p
\left(\frad{\Gamma(1-\mu)}{\Gamma(1-\nu)}\right)^{p/\mu}
\left(\frad{L}{\l_0}\right)^{p\nu/\mu}
\frad{\Gamma(1+p/\mu)\,\Gamma(1-p/\mu)}{\Gamma(1+p\nu/\mu)\,\Gamma(1-p)}.
\eeq
This result means that the scaling law of $S_L$ in Phase~IV is the following:
\beq
S_L\approx s_0\left(\frad{\Gamma(1-\mu)}{\Gamma(1-\nu)}\right)^{1/\mu}
\left(\frad{L}{\l_0}\right)^{\nu/\mu}\,Y_{\mu\nu}.
\label{s4}
\eeq
Here $Y_{\mu\nu}$ is a positive rescaled random variable
whose moment function is
\beq
m_{Y_{\mu\nu}}(p)
=\frad{\Gamma(1+p/\mu)\,\Gamma(1-p/\mu)}{\Gamma(1+p\nu/\mu)\,\Gamma(1-p)}
\qquad(-\mu<\Re p<\mu).
\label{ym4}
\eeq
The comparison of this moment function with the one of the
L\'evy variable $\LL_\mu$, equation~(\ref{xmom}), leads to the product relation
\beq
m_{Y_{\mu\nu}}(p)
=m_{\LL_\mu}(p)\,m_{\LL_\nu}(-p\nu/\mu)
=m_{\LL_\mu}(p)\,m_{{\LL_\nu}^{-\nu/\mu}}(p).
\label{pmf}
\eeq
It follows that $Y_{\mu\nu}$ can be expressed as
\beq
Y_{\mu\nu}=\LL_\mu\,{\LL_\nu}^{-\nu/\mu},
\label{y4}
\eeq
where the two independent random variables $\LL_\mu$ and $\LL_\nu$
are distributed according to L\'evy laws of respective indices $\mu$ and $\nu$.

The expression of the scaling law, equation~(\ref{s4}),
and the relation~(\ref{y4}) may also be found in a heuristic way,
that is, by eliminating $N$ between the fluctuating scaling laws~(\ref{n2})
and~(\ref{ls}).

The expression of the distribution of $Y_{\mu\nu}$ is obtained
by using~(\ref{ym4}) in the second of the formulae~(\ref{mominv}),
\beq
f_{Y_{\mu\nu}}(Y)=\int\frad{\d p}{2\pi\i}
\,\frad{\Gamma(1-p/\mu)\Gamma(1+p/\mu)}{\Gamma(1-p)\Gamma(1+p\nu/\mu)}
\,Y^{-p-1}.
\label{fy4}
\eeq
This probability density is a universal function
in the sense that it is determined by two parameters only,
namely the values of the indices $\mu$ and $\nu$.

The probability density $f_{Y_{\mu\nu}}(Y)$ can be expressed in the form of a
convergent series of powers of $Y^{\mu}$ in the case $\mu\ge\nu$ and inverse
powers of $Y^{\mu}$ in the case $\mu\le\nu$.
In the former (resp.~latter) case this series is obtained from~(\ref{fy4})
by summing the contributions of the poles of the integrand
at $p=-k\mu$ (resp.~$p=k\mu$), for $k\ge1$.
Using the difference and complement formulae for the gamma function, we find
\beq
\matrix{
f_{Y_{\mu\nu}}(Y)=\dis\sum_{k\ge1}(-1)^{k-1}\,
\frad{\sin(k\pi\nu)}{\pi}\,\frad{\Gamma(k\nu)}{\Gamma(k\mu)}\,Y^{-(1-k\mu)}
\hfill&\quad(\mu\ge\nu),\cr
f_{Y_{\mu\nu}}(Y)=\dis\sum_{k\ge1}(-1)^{k-1}\,
\frad{\sin(k\pi\mu)}{\pi}\,\frad{\Gamma(1+k\mu)}{\Gamma(1+k\nu)}\,Y^{-(1+k\mu)}
\hfill&\quad(\mu\le\nu).}
\label{sercv}
\eeq
The behaviour of the probability density at small (resp.~large) values of $Y$
is determined for all values of the indices $\mu$ and $\nu$
by the leading order term of the above expansions:
\beq
\matrix{
f_{Y_{\mu\nu}}(Y)\approx\frad{\sin\pi\nu}{\pi}\,\frad{\Gamma(\nu)}{\Gamma(\mu)}
\,Y^{-(1-\mu)}\quad\hfill&(Y\to0),\hfill\cr
f_{Y_{\mu\nu}}(Y)\approx\frad{\sin\pi\mu}{\pi}\,\frad{\Gamma(1+\mu)}
{\Gamma(1+\nu)}
\,Y^{-(1+\mu)}\quad\hfill&(Y\to\infty).\hfill}
\label{yas}
\eeq
The power law fall off of $f_{Y_{\mu\nu}}(Y)$ at large values of $Y$ is
similar to that of the L\'evy law at large values of $\LL$, as
given by~(\ref{larl}).
Its prefactor hardly depends on $\nu$, as $1/\Gamma(1+\nu)$ varies
between 1, which is reached for $\nu=0$ or $\nu=1$, and
$1/\Gamma(1+\nu_\min)=1.12917\dots$, which is reached
for the value $\nu_\min=0.46163\dots$ of $\nu$ at which the function
$\Gamma(1+\nu)$ takes its minimum $\Gamma(1+\nu_\min)=0.88560\dots$
The behaviour of $f_{Y_{\mu\nu}}(Y)$ at small values of~$Y$ is different
from that of the L\'evy law at small values of~$\LL$, as given
by~(\ref{levy0}).
It is characterized by a power law divergence in $Y^{-(1-\mu)}$,
whose amplitude is a decreasing function of $\nu$
which vanishes in the $\nu\to1$ limit.

The preceding equations take a simpler form in the particular case
of the symmetric situation where $\mu=\nu$.
Indeed, the scaling law~(\ref{s4}) has then the form
\beq
S_L\approx\frad{s_0}{\l_0}\,L\,Y_{\mu\mu}
\label{syms4}
\eeq
of a fluctuating scaling law growing proportionally to the sample length $L$.
The relation~(\ref{y4}) becomes
\beq
Y_{\mu\mu}=\frad{\LL_\mu}{\LL'_\mu}.
\label{yratio}
\eeq
Hence the random variable $Y_{\mu\mu}$ is distributed as the ratio of
two independent random variables $\LL_\mu$ and $\LL'_\mu$, each one of them
being distributed according to the same L\'evy law.
Using the difference and complement formulae for the gamma
function,~(\ref{ym4}) becomes
\beq
m_{Y_{\mu\mu}}(p)=\frad{\sin\pi p}{\mu\sin(\pi p/\mu)}
\quad(-\mu<\Re p<\mu).
\label{symym4}
\eeq
The probability density of $Y_{\mu\mu}$ admits an expression in closed form.
Indeed, each series in~(\ref{sercv}) can be summed
and both give the same result, which is
\beq
f_{Y_{\mu\mu}}(Y)=\frad{\sin\pi\mu}{\pi Y(Y^\mu+Y^{-\mu}+2\cos\pi\mu)}.
\label{lam}
\eeq
This probability density has been originally
discovered by Lamperti~\cite{lamperti}.

The overall shape of the distribution of $Y_{\mu\nu}$
is better revealed by considering the random variable
\beq
\Lambda_{\mu\nu}=\ln Y_{\mu\nu}.
\label{lamdef}
\eeq
Equation~(\ref{yas}) implies that the probability density of $\Lambda_{\mu\nu}$
falls off exponentially as $\Lambda\to\pm\infty$:
\beq
\matrix{
f_{\Lambda_{\mu\nu}}(\Lambda)\approx
\frad{\sin\pi\nu}{\pi}\,\frad{\Gamma(\nu)}{\Gamma(\mu)}
\,\e^{\mu\Lambda}\quad\hfill&(\Lambda\to-\infty),\hfill\cr
f_{\Lambda_{\mu\nu}}(\Lambda)\approx
\frad{\sin\pi\mu}{\pi}\,\frad{\Gamma(1+\mu)}{\Gamma(1+\nu)}
\,\e^{-\mu\Lambda}\quad\hfill&(\Lambda\to+\infty).\hfill}
\eeq
The moments of the distribution of $\Lambda_{\mu\nu}$
may be calculated by taking advantage of the identity
$\w f_{\Lambda_{\mu\nu}}(p)=m_{Y_{\mu\nu}}(-p)$.
Expanding the moment function of~$Y_{\mu\nu}$,
equation~(\ref{ym4}), in powers of $p$, we find
\beq
\mean{\Lambda_{\mu\nu}}=\frad{(\nu-\mu)\,{\bf C}}{\mu},\quad
\cum{\Lambda_{\mu\nu}^2}=\frad{(2-\mu^2-\nu^2)\pi^2}{6\mu^2},
\label{lmom}
\eeq
where ${\bf C}=0.57721\dots$ is the Euler-Mascheroni constant, and so on.
The first of these relations indicates that the mean of the
distribution is positive (resp.~negative)
if $\mu<\nu$ (resp.~$\mu>\nu$), the second that its variance
decreases as the values of one or both indices increase.
Mean and variance vanish as $\mu\to1$ and $\nu\to1$ simultaneously,
in agreement with the limiting result~(\ref{y11}).

If $\mu=\nu$, a closed form expression of the probability density
of $\Lambda_{\mu\mu}=\ln Y_{\mu\mu}$ is easily obtained from~(\ref{lam}).
We find the symmetric law
\beq
f_{\Lambda_{\mu\mu}}(\Lambda)
=\frad{\sin\pi\mu}{2\pi(\cosh\mu\Lambda+\cos\pi\mu)}.
\label{Lam}
\eeq

\begin{figure}[htb]
\begin{center}
\includegraphics[angle=90,height=7truecm]{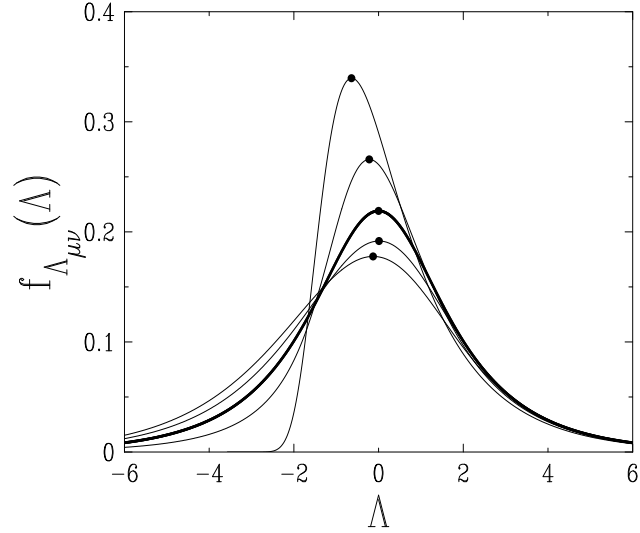}
\caption{\small
Graph of the probability density of the random variable
$\Lambda_{\mu\nu}=\ln Y_{\mu\nu}$
for $\mu=0.6$ and five different values of $\nu$.
From top to bottom near $\Lambda=0$: $\nu=1$
(case in which $Y_{\mu1}=\LL_\mu$ is distributed
according to the L\'evy law~(\ref{flevy})),
$\nu=0.8$, $\nu=0.6$ (thick line, case
of the symmetric probability density~(\ref{Lam})), $\nu=0.4$, and $\nu=0.2$.
A heavy dot indicates the maximum of the probability density
for each value of $\nu$.}
\label{fdensity}
\end{center}
\end{figure}

The probability density $f_{\Lambda_{\mu\nu}}(\Lambda)$ is represented in
Figure~\ref{fdensity} for $\mu=0.6$ and five different values of $\nu$.
The curves result from a numerical integration of the following expression:
\beq
f_{\Lambda_{\mu\nu}}(\Lambda)=\frad{1}{2\mu}\int_{-\infty}^\infty
\frad{y\,\e^{-\i y\Lambda}}
{\sinh(\pi y/\mu)\Gamma(1-\i y)\Gamma(1+\i y\nu/\mu)}\,\d y.
\eeq
This integral representation of the probability density is obtained
by setting $p=\i y$ in~(\ref{fy4}) and using the relation
$\Gamma(1+\i u)\Gamma(1-\i u)=\pi u/\sinh(\pi u)$.
The figure confirms that the shape of $f_{\Lambda_{\mu\nu}}(\Lambda)$
is very sensitive to the value of $\nu$ on the negative side,
but much less so on the positive side.
The position of its maximum (shown by a heavy dot)
moves only weakly as a function of $\nu$, and in a manner that is not monotonic.

The scaling laws of the first three
phases can be obtained from that of Phase~IV by taking the right
limits and using the proper identifications.

\begin{itemize}

\item For $\mu\to1$,
the results of Phase~III are recovered with the identification
\beq
\mean{s}=\lim_{\mu\to1}\frad{s_0}{1-\mu}.
\label{id3}
\eeq
In particular, $\LL_\mu\to1$, and so~(\ref{y4}) becomes
\beq
Y_{1\nu}=Z_\nu=\LL_\nu^{-\nu},
\eeq
in agreement with~(\ref{zxi}).

\item For $\nu\to1$,
the results of Phase~II are recovered with the identification
\beq
\mean{\l}=\lim_{\nu\to1}\frad{\l_0}{1-\nu}.
\label{id2}
\eeq
In particular, $\LL_\nu\to1$, and so~(\ref{y4}) becomes
\beq
Y_{\mu1}=\LL_\mu,
\label{ymu1}
\eeq
in agreement with~(\ref{l2}).

\item For $\mu\to1$ and $\nu\to1$ simultaneously,
the results of Phase~I are recovered
with the identifications~(\ref{id3}) and~(\ref{id2}).
In particular,~(\ref{y4}) becomes
\beq
Y_{11}=1.
\label{y11}
\eeq
The rescaled random variable becomes deterministic in this limiting case,
which implies that $S_L$ becomes self-averaging, in agreement with~(\ref{l1}).

\end{itemize}

\begin{figure}[htb]
\begin{center}
\includegraphics[angle=90,height=7truecm]{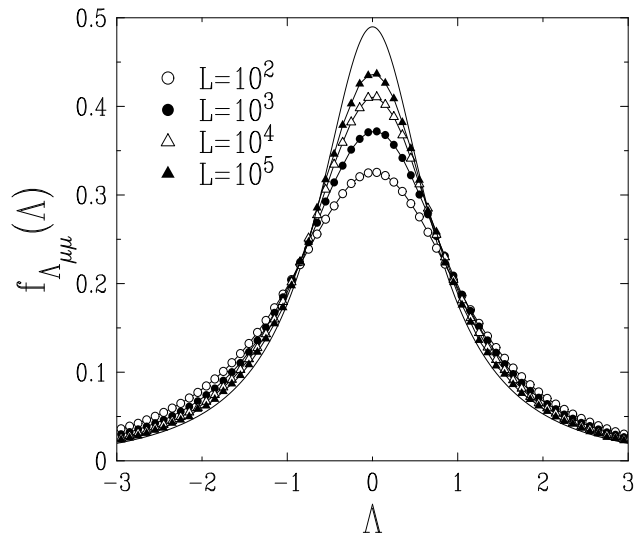}
\caption{\small
Study of the convergence of the distribution of
$S_L$ towards the asymptotic scaling law~(\ref{s4}).
Symbols: histograms of numerical data for $\ln(S_L/L)$
for $\mu=\nu=0.8$ and four values of the sample length~$L$.
Solid line: theoretical prediction for the asymptotic distribution
of $\Lambda_{\mu\mu}$, equation~(\ref{Lam}).}
\label{fhistos}
\end{center}
\end{figure}

Finally, we have addressed by means of numerical simulations
the question of the convergence of the distribution of
$S_L$ for a finite sample length $L$ towards the asymptotic
scaling law~(\ref{s4}).
We have generated ensembles of $10^8$ samples for
four values of $L$ ranging from $10^2$ to $10^5$.
Each sample consisted of a sequence of pseudo-random
values of $\l_n$ and $s_n$, distributed according to the truncated power laws
\beq
f_s(s)=\frad{\mu}{s^{1+\mu}}\quad(s>1),\qquad
f_\l(\l)=\frad{\nu}{\l^{1+\nu}}\quad(\l>1).
\eeq
These distributions have $s_0=\l_0=1$.
For definiteness, we have considered the symmetric case $\mu=\nu$.
In this case equation~(\ref{s4}) becomes~(\ref{syms4}) which implies
\beq
\ln\frac{S_L}{L}\approx\Lambda_{\mu\mu}.
\eeq
The choice of the rather high value $\mu=\nu=0.8$ for the index was motivated
by the expectation that finite-size effects are larger for larger $\mu$
because of the presence of logarithmic corrections to scaling for $\mu=1$.
An example of such corrections is mentioned at the end of Section~5.
Figure~\ref{fhistos} shows histogram plots of our numerical data
for $\ln(S_L/L)$ for the four values of the sample length $L$.
A slow and monotonic convergence of the data
towards the asymptotic prediction~(\ref{Lam}) is observed.

\section{Discussion}

In the present work we have studied a model
of quantum transport in one dimension
based on a disordered array of units whose lengths $\l_n$ and actions $s_n$
are independent random variables with either narrow or broad distributions.
We have successively investigated the statistical ensemble
at fixed number $N$ of units (in Section~3)
and the statistical ensemble at fixed sample length $L$ (in Section~4).
The model is richer in the latter ensemble than in the former
because the sample action $S_L$ is the sum of a random rather than fixed number
$N_L$ of independent random variables.
Equation~(\ref{mw}) has been the starting point of the analysis
in the ensemble at fixed sample length.
This equation is analogous to the Montroll-Weiss equation~\cite{mw},
which is well-known in the theory of the continuous time random walks.
Nevertheless the overlap between the present work and the abundant literature
on continuous time random walks (see, e.g.,~\cite{ctrw}),
and more generally on fractional diffusion (see, e.g.,~\cite{fd}),
seems to be minute.
This is essentially because our basic random variables (lengths and actions)
are intrinsically positive
whereas most works on fractional diffusion consider displacements
which are symmetrically distributed random variables,
and so all the results differ.
Let us however mention that the distribution of our variable $Y_{\mu\nu}$
appears in some form in Reference~\cite{sz}.

\begin{figure}[htb]
\begin{center}
\includegraphics[angle=90,height=6truecm]{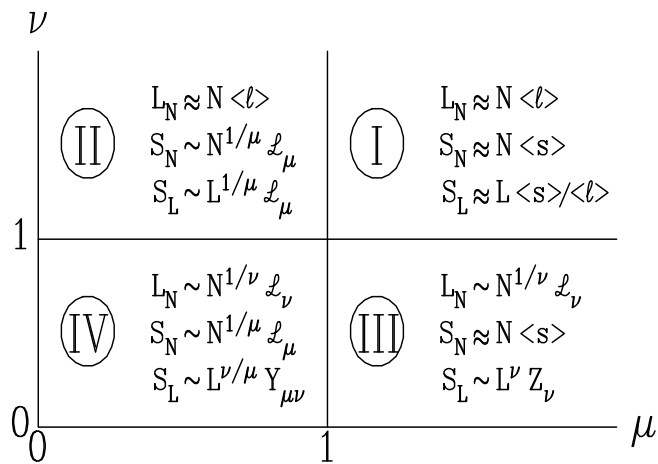}
\caption{\small
Phase diagram of the model in the plane of the indices $\mu$ and $\nu$,
showing the four phases labelled I to IV.
The scaling behaviour of $S_N$ and $L_N$ at fixed $N$
(Section~3) and of $S_L$ at fixed $L$ (Section~4) is given for each phase.}
\label{fdiagramme}
\end{center}
\end{figure}

The main outcomes of this work, that is,
those concerning the ensemble at fixed sample length derived in Section~4,
are summarized in the phase diagram shown in Figure~\ref{fdiagramme}.
Four different phases can be distinguished,
depending on whether the indices~$\mu$ and $\nu$ are greater or less than unity.
A fluctuating scaling behaviour for the action $S_L$, of the form
\beq
S_L\approx a\,Y\,L^\alpha,
\label{fsd}
\eeq
has been shown to hold in Phases~II, III, and~IV.
This fluctuating scaling law consists of three factors:
a power law $L^\alpha$ of the sample length,
with a scaling exponent $\alpha$ whose value
dictates the localization properties of the eigenstates, as detailed below;
a rescaled random variable $Y$, which has a non-trivial
asymptotic distribution with some universal character;
and a non-universal prefactor~$a$,
which encompasses the microscopic details of the model.
Table~\ref{tab} gives the expressions of the exponent $\alpha$
and of the rescaled random variable $Y$ in the four phases of the model.
Remarkably enough, $Y$ can always be expressed
in terms of positive L\'evy variables $\LL_\mu$.

\begin{table}[ht]
\begin{center}
\begin{tabular}{|c|c|c|c|}
\hline
Phase&Indices&$\alpha$&$Y$\\
\hline
I&$\mu>1$,~$\nu>1$&1&$1$\\
\hline
II&$\mu<1$,~$\nu>1$&$1/\mu$&$\LL_\mu$\\
III&$\mu>1$,~$\nu<1$&$\nu$&$Z_\nu={\LL_\nu}^{-\nu}$\\
IV&$\mu<1$,~$\nu<1$&$\nu/\mu$&$Y_{\mu\nu}=\LL_\mu\,{\LL_\nu}^{-\nu/\mu}$\\
\hline
\end{tabular}
\end{center}
\caption{Expressions of the exponent $\alpha$
and of the rescaled random variable $Y$
involved in the fluctuating scaling law~(\ref{fsd})
for the sample action $S_L$ in the four phases of the model.
Phase~I: $Y=1$ is trivial, and so $S_L$ is self-averaging.
Phases~II, III, and~IV: $Y$ is non-trivial,
and so $S_L$ is not self-averaging.}
\label{tab}
\end{table}

The physical consequences of the fluctuating scaling law~(\ref{fsd})
are the following.
As recalled in the introduction,
in the usual one-dimensional localization problem,
where the random potential has short-range correlations,
the conductance, which is proportional
to the transmission probability $\abs{\TT}^2$,
is a strongly fluctuating quantity
in the insulating regime of long enough samples ($L\gg\xi$).
On the other hand, the action $S$
is much better behaved, as it is self-averaging in a strong sense.
Whenever the fluctuating scaling law~(\ref{fsd}) holds,
the action itself is not self-averaging
but keeps on fluctuating in samples whose length is arbitrarily large.
The same feature holds for the sample-dependent effective localization length,
defined as
\beq
\xi_L=\frad{L}{S_L},
\label{xidef}
\eeq
for which~(\ref{fsd}) implies
\beq
\xi_L\approx\frad{L^{1-\alpha}}{a\,Y}.
\label{xieff}
\eeq

The lack of self-averaging of the action $S_L$ and of the effective
localization length~$\xi_L$,
which occurs in three of the four phases of the model,
is one of its most salient features.
As recalled in the introduction, the lack of self-averaging of the action
was also observed in the superlocalized regime
of one-dimensional models with non-stationary random potentials
whose fluctuations grow with distance~\cite{super1,super2,jmls}.
We are therefore tempted to argue that a fluctuating scaling behaviour
of the form~(\ref{fsd}) is a generic feature of quantum transport
in a large universality class of one-dimensional systems with broad disorder,
in the sense of random potentials with strong
and long-range correlated fluctuations.

Three different regimes of localization are allowed by the
fluctuating scaling law~(\ref{fsd}),
according to the value of the scaling exponent $\alpha$.

\begin{itemize}

\item $\alpha<1$
(Phase~IV for $\mu>\nu$ and Phase~III): {\it Underlocalization.}
In this regime the typical sample action $S_L$ grows less rapidly
than linearly with the sample length $L$.
The effective localization length therefore grows with $L$
as the power law $\xi_L\sim L^{1-\alpha}$.
This underlocalized regime is intermediate between localized and extended.
It is qualitatively reminiscent of the behaviour right at the critical point
corresponding to the metal-insulator transition in higher dimension.

\item $\alpha>1$
(Phase~IV for $\mu<\nu$ and Phase~II): {\it Superlocalization.}
In this regime the typical sample action $S_L$ grows more rapidly
than linearly with the sample length~$L$.
The effective localization length therefore decreases with
$L$ as the power law $\xi_L\sim L^{-(\alpha-1)}$.
The positive exponent $\alpha-1$ can be arbitrarily large if $\mu\ll1$.
This superlocalized regime is similar to that observed
in one-dimensional models with non-stationary
random potentials~\cite{super1,super2,jmls}.
For such potentials, however, we have $\alpha=1+H/2$,
with the Hurst exponent of the random potential satisfying $0<H<1$,
and so the value of $\alpha$ cannot be greater than $3/2$.

\item $\alpha=1$
(Phase~IV for $\mu=\nu$): {\it Fluctuating localization.}
In this regime the typical sample action~$S_L$ grows linearly with $L$,
just as in the usual one-dimensional localization problem,
where the random potential has short-range correlations.
However, at variance with the usual situation,
the action keeps on fluctuating in samples whose length is arbitrarily large.
The effective localization length $\xi_L$
also keeps on fluctuating from sample to sample,
without ever systematically increasing or decreasing
as the sample length is increased.
Its probability distribution is indeed asymptotically independent
of the sample length.
The term of fluctuating localization is therefore fully justified.
Equation~(\ref{syms4}) leads to the expression
\beq
\xi_L\approx\frad{\l_0}{s_0 Y_{\mu\mu}},
\eeq
where $Y_{\mu\mu}$ is distributed
as the ratio of two L\'evy variables with the same index,
equation~(\ref{yratio}).
Therefore, apart from the prefactor $\l_0/s_0$,
the effective localization length $\xi_L$ is also
distributed as the ratio of two L\'evy variables with the same index~$\mu$,
and so its probability density is also given by a Lamperti law~(\ref{lam}).

\end{itemize}

Along the borderlines which separate the various phases
shown in Figure~\ref{fdiagramme},
that is, when at least one of the indices is equal to unity,
the above results are affected by logarithmic corrections to scaling.
The simplest case occurs in the ensemble at fixed number $N$ of units.
The approach of Section~3 leads
to the following result in the marginal case $\mu=1$:
\beq
S_N\approx N\,s_0\,(\ln N+C+\eta).
\label{nmar}
\eeq
Here $C$ is a non-universal constant which depends
on the whole shape of the probability distribution $f_s(s)$
and $\eta$ is a rescaled random variable
whose universal probability distribution is
characterized by the following moment generating function:
\beq
\w f_\eta(y)=\e^{y\ln y}.
\eeq
The sample action $S_N$ may be called marginally self-averaging,
as its mean grows as $N\ln N$ whereas its fluctuations grow as $N$.

\appendix
\setcounter{equation}{0}
\def\theequation{A.\arabic{equation}}
\def\thesubsection{A.\arabic{subsection}}
\section*{Appendix.~Scattering and transport in the weak transmission regime}

This appendix is devoted to the study of the scattering and transport properties
of a single unit and of an array of $N$ units.
We first recall the transfer matrix formalism
for a single unit~\cite{pen,cpv,alea}.
We then use this formalism to obtain
the amplitude of transmission through an array of $N$ units
in the regime where the transmission through each unit is small.
We derive the law of addition for the actions~(\ref{adds}),
which is the starting point of the model investigated in this paper.

\subsection{Transfer matrix formalism for a single unit}

Let us first consider the scattering of a quantum particle by a single unit.
The time-independent Schr\"odinger equation corresponding to this process is
\beq
-\psi''(x)+V(x)\psi(x)=E\psi(x),
\label{sch}
\eeq
where the potential $V(x)$ has an arbitrary form in the scattering unit
($a\le x\le b$) and vanishes elsewhere.
In both half-lines (perfect leads) on either side of the unit,
the wavefunction is a superposition of plane waves $\exp(\pm\i kx)$,
with $k=\dis{\sqrt{E}}$:
\beq
\psi(x)=\left\{\matrix{
A\,\e^{\i kx}+B\,\e^{-\i kx}\quad\hfill&{\rm on~the~left~of~the~unit}\hfill&
(x<a),\hfill\cr
C\,\e^{\i kx}+D\,\e^{-\i kx}\hfill&{\rm on~the~right~of~the~unit}\hfill&(x>b).
\hfill}\right.
\label{wave}
\eeq
The amplitudes $A$, $B$, $C$, and $D$ are linked together by linear relations
of the form
\beq
\pmatrix{C\cr D}=M\pmatrix{A\cr B},
\eeq
where $M$ is the transfer matrix of the unit.
Time reversal invariance implies that $M$ can be parametrized as
\beq
M=\pmatrix{e&f\cr f^\star&e^\star},
\label{fg}
\eeq
where the star denotes complex conjugation.
In addition, conservation of the probability current leads to the condition that
\beq
\det M=\abs{e}^2-\abs{f}^2=1.
\eeq

Let $r$ (resp.~$r'$) and $t$ (resp.~$t'$) denote the reflection and
transmission amplitudes for a particle coming from the left (resp.~right).
The first case corresponds to the values $A=1$, $B=r$, $C=t$, and $D=0$,
the second to the values $A=0$, $B=t'$, $C=r'$, and $D=1$.
Thus we have
\beq
t=t'=\frad{1}{e^\star},\quad r=-\frad{f^\star}{e^\star},
\quad r'=\frad{f}{e^\star}.
\label{trtr}
\eeq
These quantities satisfy the relations
\beq
\abs{r}^2=\abs{r'}^2=1-\abs{t}^2,\quad r't^\star+r^\star t=0.
\label{rels}
\eeq
The above relations allow one to derive several equivalent parametrizations
of the transfer matrix in terms of the reflection and transmission amplitudes;
the one we choose here is the following:
\beq
M=\pmatrix{1/t^\star&-r^\star/t^\star\cr-r/t&1/t}.
\label{tm}
\eeq

\subsection{Many units in the weak transmission regime}

Let us now consider the transmission through a sample consisting of an
arbitrary number~$N$ of units which are put end to end.
The transfer matrix $\MM_N$ of the whole sample
is the ordered product of the transfer matrices $M_n$ of the units,
\beq
\MM_N=M_N\cdots M_1.
\label{tproduct}
\eeq
This equation is equivalent to the matrix recursion equation
\beq
\MM_N=M_N\MM_{N-1}.
\label{prod}
\eeq

By analogy with the expression~(\ref{tm}) of the transfer matrix for a
single unit, we parametrize $\MM_N$ as
\beq
\MM_N=\pmatrix{1/\TT_N^\star&-\RR_N^\star/\TT_N^\star\cr-\RR_N/\TT_N&1/\TT_N},
\label{ftt}
\eeq
where $\RR_N$ and $\TT_N$ are the reflection and transmission amplitudes
corresponding to the whole sample.
Equation~(\ref{prod}) leads to the following non-linear recursion relations:
\beq
\frad{1}{\TT_N}=\frad{1}{t_N\TT_{N-1}}+\frad{r_N\RR_{N-1}^\star}
{t_N\TT_{N-1}^\star},\quad
\frad{\RR_N}{\TT_N}
=\frad{r_N}{t_N\TT_{N-1}^\star}+\frad{\RR_{N-1}}{t_N\TT_{N-1}},
\label{recur}
\eeq
with the initial conditions $\TT_0=1$ and $\RR_0=0$.
Combining these two relations, we obtain
a three-term non-linear recursion relation linking $\TT_N$
to $\TT_{N-1}$ and $\TT_{N-2}$, which is
\beq
\frad{1}{\TT_N}=\frad{1}{t_N\TT_{N-1}}+\frad{t_{N-1}r_N}{r_{N-1}t_N}
\left(\frad{1}{{t_{N-1}^\star}\TT_{N-1}}-\frad{1}{\TT_{N-2}}\right).
\label{com}
\eeq
Setting
\beq
\TT_N=\frac{t_1\dots t_N}{\DD_N}
\eeq
and using the second relation of~(\ref{rels}),
we find that $\DD_N$ is linked to $\DD_{N-1}$ and $\DD_{N-2}$ by
the following linear recursion relation:
\beq
\DD_N=\DD_{N-1}-\frad{r'_{N-1}r_N}{1-{\vert t_{N-1}\vert}^2}
\left(\DD_{N-1}-{\vert t_{N-1}\vert}^2\DD_{N-2}\right).
\label{deter}
\eeq

We are interested here in the expression of $\TT_N$ in the regime in which
the probability of transmission through each unit is small.
In this regime, neglecting $\abs{t_{N-1}}^2$ with respect to unity,
(\ref{deter}) simplifies to
\beq
\DD_N\approx(1-r'_{N-1}r_N)\DD_{N-1},
\label{mdet}
\eeq
with $\DD_1=1$.
Iterating this equation, we obtain
\beq
\DD_N\approx\prod_{n=1}^{N-1}(1-r'_nr_{n+1}).
\label{appro}
\eeq
We are thus left with the following expression for the transmission amplitude
$\TT_N$ in the weak transmission regime:
\beq
\TT_N\approx\frad{\prod_{n=1}^N t_n}{\prod_{n=1}^{N-1}(1-r'_nr_{n+1})}.
\label{otr}
\eeq
This leading order result is valid up to corrections
which are proportional to the $\abs{t_n}^2$ in relative value
and could be derived from the full recursion relation~(\ref{deter}).
The result~(\ref{otr}) can be recast in the following more appealing way:
\beq
\TT_N\approx
t_1\,\frad{1}{1-r'_1r_2}\,t_2\,\frad{1}{1-r'_2r_3}
\,t_3\dots t_{N-1}\,\frad{1}{1-r'_{N-1}r_N}\,t_N.
\eeq
Here each denominator corresponds to the geometric resummation
of all the repeated reflections between two consecutive units.
These are indeed the only internal reflections from the whole
multiple scattering expansion that survive at leading order
in the weak transmission regime.

The leading order expression of the sample action is therefore
\beq
S_N=-\half\ln\abs{\TT_N}^2
\approx-\half\sum_{n=1}^N\ln\abs{t_n}^2+\Re\sum_{n=1}^{N-1}\ln(1-r'_nr_{n+1}),
\eeq
i.e.,
\beq
S_N\approx\sum_{n=1}^Ns_n+\Re\sum_{n=1}^{N-1}\ln\left(1-{\abs{r_n}}{\abs{r_{n+1}}}
\,\e^{\i(\theta'_n+\theta_{n+1})}\right),
\label{rag}
\eeq
where we have introduced the action $s_n$
and the reflection angles $\theta_n$ and $\theta'_n$ of each unit,
according to
\beq
s_n=-\half\ln\abs{t_n}^2,\quad
r_n=\abs{r_n}\e^{\i\theta_n},\quad r'_n=\abs{r_n}\e^{\i\theta'_n}.
\eeq

We now introduce our second hypothesis besides the weak transmission regime,
namely that the internal reflection angles $\theta_n$ and $\theta'_n$
are random and uniformly distributed between $0$ and $2\pi$.
This hypothesis can be postulated as a mere simplifying assumption,
in keeping with a long tradition
in localization theory~\cite{lan,pen,four,abst}.
It can alternatively be justified on physical grounds
by considering that the reflection angles are rapidly varying functions
of the incoming momentum $k$.
This is especially true in the weak transmission regime,
in which the length $\l_n$ of each unit is expected
to be such that $k\l_n\gg1$.
As a consequence, any narrow distribution of incoming momentum will result
in a uniform averaging over the reflection angles.
Carrying out this averaging over the angles $\theta_n$
and $\theta'_n$ in~(\ref{rag}), we
find that the second sum vanishes by virtue of the identity
\beq
\int_0^{2\pi}\ln(1-z\,\e^{i\theta})\,\d\theta
=-\sum_{k\ge1}\frad{z^k}{k}
\underbrace{\int_0^{2\pi}\e^{ik\theta}\,\d\theta}
_{\dis{2\pi\,\delta_{k0}}}=0,
\eeq
which holds for any number $z$ such that $\abs{z}\le1$.

We thus find that the action $S_N$ of the whole sample is approximately
given by
\beq
S_N\approx\sum_{n=1}^N s_n.
\label{adds}
\eeq
This simple formula,
which can be referred to as the {\it law of addition for the actions},
is the starting point of the model investigated in this paper.


\Bibliography{99}

\bibitem{l1}
Lifshitz I M, Gredeskul S A and Pastur L A 1988 {\it Introduction to
the Theory of Disordered Systems} (New York: Wiley)

\bibitem{l2}
Kramer B and MacKinnon A 1993 {\it Rep. Prog. Phys.} {\bf 56} 1469

\bibitem{pen}
Pendry J B 1994 {\it Adv. Phys.} {\bf 43} 461

\bibitem{lan}
Landauer R 1970 {\it Phil. Mag.} {\bf 21} 863

\bibitem{il}
Imry Y and Landauer R 1999 {\it Rev. Mod. Phys.} {\bf 71} S306 and references
therein

\bibitem{four}
Anderson P W, Thouless D J, Abrahams E and Fisher D S 1980 {\it Phys. Rev. B}
{\bf 22} 3519

\bibitem{aaa}
Abrikosov A A 1981 {\it Solid State Commun.} {\bf 37} 997

\bibitem{cpv}
Crisanti A, Paladin G and Vulpiani A 1992 {\it Products of Random Matrices in
Statistical Physics} (Berlin: Springer)

\bibitem{alea}
Luck J M 1992 {\it Syst\`emes d\'esordonn\'es unidimensionnels
(Collection Al\'ea-Saclay)}

\bibitem{super1}
de Moura F A B F and Lyra M L 1998 {\it Phys. Rev. Lett.} {\bf 81} 3735
\nonum
\dash\ 1999 {\it Physica A} {\bf 266} 465
\nonum
\dash\ 2000 {\it Phys. Rev. Lett.} {\bf 84} 199

\bibitem{super2}
Kantelhardt J W, Russ S, Bunde A, Havlin S and Webman I 2000 {\it Phys. Rev.
Lett.} {\bf 84} 198
\nonum
Bunde A, Havlin S, Kantelhardt J W, Russ S and Webman I 2000 {\it J. Mol. Liq.}
{\bf 86} 151
\nonum
Russ S, Kantelhardt J W, Bunde A and Havlin S 2001 {\it Phys. Rev. B} {\bf 64}
134209

\bibitem{jmls}
Luck J M 2005 {\it J. Phys. A} {\bf 38} 987

\bibitem{renewal}
Cox D R 1962 {\it Renewal Theory} (London: Methuen)
\nonum
Cox D R and Miller H D 1965 {\it The Theory of Stochastic Processes}
(London: Chapman~\& Hall)

\bibitem{gry}
Gradshteyn I S and Ryzhik I M 1965 {\it Table of Integrals, Series, and
Products} (New York and London: Academic Press)

\bibitem{gl}
Godr\`eche C and Luck J M 2001 {\it J. Stat. Phys.} {\bf 104} 489

\bibitem{levy1}
L\'evy P 1954 {\it Th\'eorie de l'addition des variables al\'eatoires}
(Paris: Gauthier-Villars)
\nonum
Gnedenko B V and Kolmogorov A N 1954 {\it Limit Distributions for Sums of
Independent Random Variables} (Reading MA: Addison-Wesley)

\bibitem{levy2}
Bouchaud J P and Georges A 1990 {\it Phys. Rep.} {\bf 195} 127
\nonum
Shlesinger M F, Zaslavsky M G and Frisch U 1995 {\it L\'evy Flights
and Related Topics in Physics} Lecture Notes in Physics {\bf 450}
(Berlin: Springer)

\bibitem{mw}
Montroll E W and Weiss G H 1965 {\it J. Math. Phys.} {\bf 6} 167

\bibitem{ctrw}
Hughes B D 1995 {\it Random Walks and Random Environments Volume 1:
Random Walks} (Oxford: Oxford University Press)

\bibitem{lamperti}
Lamperti J 1958 {\it Trans. Amer. Math. Soc.} {\bf 88} 380

\bibitem{fd}
Metzler R and Klafter J 2000 {\it Phys. Rep.} {\bf 339} 1
\nonum
\dash\ 2004 {\it J. Phys. A} {\bf 37} R161

\bibitem{sz}
Saichev A I and Zaslavsky G M 1997 {\it Chaos} {\bf 7} 753

\bibitem{abst}
Abrahams E and Stephen M 1980 {\it J. Phys. C} {\bf 13} L377

\end{thebibliography}
\end{document}